\documentclass[twocolumn,tighten]{aastex63}
\hypersetup{linkcolor=red,citecolor=blue,filecolor=cyan,urlcolor=magenta}

\usepackage[fleqn]{amsmath}
\usepackage{amssymb}
\usepackage{txfonts}
\usepackage{graphicx}
\usepackage{multirow}
\usepackage{dcolumn}
\usepackage{array}
\usepackage{xcolor}
\usepackage{float}

\received{October 24, 2019}
\revised{December 6, 2019}
\submitjournal{AAS}
\shorttitle{Albedo, Temperature and Active Mechanism of 133P}
\shortauthors{Yu, Hsia \& Ip et al.}

\begin{document}

\title{Low activity main belt comet 133P/Elst-Pizarro: New constraints on its Albedo, Temperature and Active Mechanism from a thermophysical perspective}

\correspondingauthor{Liang Liang Yu}
\email{yullmoon@live.com}

\author{Liang Liang Yu}
\affiliation{State Key Laboratory of Lunar and Planetary Sciences, Macau University of Science and Technology, Macau, China}

\author{Chih Hao Hsia}
\affiliation{State Key Laboratory of Lunar and Planetary Sciences, Macau University of Science and Technology, Macau, China}

\author{Wing-Huen Ip}
\affiliation{State Key Laboratory of Lunar and Planetary Sciences, Macau University of Science and Technology, Macau, China}
\affiliation{Institute of Astronomy, National Central University, Jhongli, Taoyuan City 32001, Taiwan}

\begin{abstract}
133P/Elst-Pizarro is the firstly recognized main-belt comet, but we still know little about its nucleus. Firstly we use mid-infrared data of Spitzer-MIPS, Spitzer-IRS and WISE to estimate its effective diameter $D_{\rm eff}=3.9^{+0.4}_{-0.3}$ km, geometric albedo $p_{\rm v}=0.074\pm0.013$ and mean Bond albedo $A_{\rm eff,B}=0.024\pm0.004$. The albedo is used to compute 133P's temperature distribution, which shows significant seasonal variation, especially polar regions, ranging from $\sim40$ to $\sim200$ K. Based on current activity observations, the maximum water gas production rate is estimated to be $\sim1.4\times10^{23}\rm~s^{-1}$, being far weaker than $\sim10^{26}\rm~s^{-1}$ of JFC 67P at similar helio-centric distance $\sim2.7$ AU, indicating a thick dust mantle on the surface to lower down the gas production rate. The diameter of the sublimation area may be $<\sim200$ m according to our model prediction. We thus propose that 133P's activity is more likely to be caused by sublimation of regional near-surface ice patch rather than homogeneous buried ice layer. Such small near-surface ice patch might be exposed by one impact event, before which 133P may be an extinct comet (or ice-rich asteroid) with ice layer buried below $\sim$40 m depth. The proposed ice patch may be located somewhere within latitude $-50\sim50^\circ$ by comparing theoretical variation of sublimation temperature to the constraints from observations. The time scale to form such a thick dust mantle is estimated to be $>100$ Myr, indicating that 133P may be more likely to be a relatively old planetesimals or a member of an old family than a recently formed fragment of some young family.
\end{abstract}


\section{Introduction}
133P/Elst-Pizarro (hereafter 133P) was originally discovered as an asteroid-like
point source with no special characteristics in the main belt by the Siding Spring
1.2-m telescope on July 24, 1979, when it was at mean anomaly $\sim15.3^\circ$,
thus being named as asteroid 1979 OW7 \citep{McNaught1996}. \citet{McNaught1996}
also reported that this object was still a point-source on September 15, 1985, when
it was at mean anomaly $\sim48.4^\circ$. Then on August 7, 1996, Eric W. Elst and
Guido Pizarro observed a main-belt object showing a long narrow dust tail but no gas
feature from the ESO 1-m Schmidt telescope at La Silla Observatory. This special object
looked like a comet, and thus was designated as comet P/1996 N2, which turned out to be
the already discovered main-belt asteroid 1979 OW7. Subsequently the object obtained its
current name 133P/Elst-Pizarro.

The phenomenon that 133P suddenly showed comet-like dust tail but no observable gas coma or
gas tail is quite strange for a main-belt object with the Tisserand parameter $T_J=3.18>3$,
because typical comets like Jupiter Family Comets (JFCs) as well as Halley-family comets (HFCs)
have $T_J<3$. If the observation in 1979 and 1985 did tell us that 133P was inactive
asteroid at that time, then the activity observed on August 7, 1996, when 133P was at mean
anomaly $\sim25.2^\circ$, seems to be triggered suddenly at some particular time between
1985 and 1996. For instance, \citet{Toth2000} proposed that the dust tail of 133P was caused
by a recent impact event, which could disturb the surface and generate ejection of surface
dust material.

\citet{Hsieh2004} reported the recurrent dust activity of 133P in its 2002 perihelion passage,
which lasted at least 5 months from 2002 August to December based on observations by the UH
2.2-m telescope in 2002 and the Keck I 10-m telescope in 2003, the hypotheses of dust ejection
by one-time impact event to explain the appearance of 133P's comet-like tail in 1996 was thus
ruled out. \citet{Hsieh2004} considered a variety of mechanisms to explain the observed comet-like
behavior of 133P, but preferred to explain the dust tail of 133P to be the result of seasonal
sublimation of exposed surface ice, raising the interesting question about when 133P would be
comet-like active and when it would be inactive along its orbit.

For this purpose, \citet{Hsieh2010} carried out a multi-year monitoring campaign of 133P
from 2003 to 2008 (nearly an orbital cycle of 133P), and again observed the return of
its activity in 2007. They found that 133P looks like an asteroid at most part of its
orbit, but can also display dust tail feature like a comet when it was close to or shortly
after perihelion in 1996, 2002, and 2007. Moreover, the recurrence of dust-tail activity
of 133P near perihelion was also observed on July 10, 2013 by the {\it Hubble Space Telescope}
\citep{Jewitt2014}. Such significant seasonal variation and cyclical recurring activity
strongly support the idea that the dust ejection activity is caused by ice sublimation, and
further imply that there should even exist groups of icy small bodies in the main belt, which
led to the discovery of a new comet group, named "Main-Belt Comet" \citep[MBC,][]{Hsieh2006}.

\citet{Hsieh2004,Hsieh2010} tended to explain the recurring activity of 133P by seasonal
sublimation of regional surface icy patch, which may be exposed by impacts from deeply
buried icy layer. This model seemed to be perfect at that time. However, following the
discovery of more and more main-belt comets, \citet{Hsieh2015} found that nearly all of
the known MBCs, appeared to show activity close to or shortly after perihelion. If sublimation
of regional surface icy patch is responsible for these observed activity, there is no reason
to expect all the exposed icy patch on these MBCs to get local summer close to or shortly
after perihelion, because impacts on the surface should be random events. Therefore,
\citet{Hsieh2015} proposed another possible mechanism. That is variation of sublimate rate
of homogenous buried icy layer due to change of heliocentric distance may be the cause.
A new question thus arises on whether activities of MBCs are caused by sublimation of
regional surface ice patches or by sublimation of homogenous buried ice layer?

On the other hand, the discovery of MBCs implies that water ice can survive in the main belt
even today since their formation. Details of the physical properties of the MBC nuclei can
give us key information about the formation and evolution of the main belt, and hence provide
clues about the formation and evolution of the solar system. Clarification of this issue would
also shed light on the origin of water on terrestrial planets like our Earth. However, distances
to MBCs are too far away for current telescopes to figure out what happens on such MBC nuclei.
So spacecraft mission to MBCs would be necessary and meaningful. This is the reason why 133P
becomes the target of a proposed ESA spacecraft mission named 'Castalia' \citep{Snodgrass2018},
and it was also selected to be a target of a proposed Chinese small-body mission. Thus theoretical
modelling and constraints about the thermal environment and thermal activity prior to the space
mission would be of significance for both the mission planning and instruments design.

In this paper, we aim to figure out the active mechanism of 133P, and estimate its albedo,
temperature and gas/dust production rate. To realize these goals, firstly we use the radiometric
method to infer the albedo and size of the nucleus of 133P, then simulate the possible temperature
variation of the surface layers based on the estimated albedo and thermal parameters. Finally
dust-ice two-layer sublimation model of buried ice is utilized to explain the current available
observations on the activity of 133P, which enables us to depict the possible distribution of
ice on 133P and orientation of 133P's rotation axis as well. The results show that the activity
of 133P is more likely to be caused by the sublimation of exposed regional near-surface ice patches
than homogenous buried icy layer.

\section{Radiometric Constraints}
\subsection{Thermal Infrared observations}
\subsubsection{Spitzer MIPS data}
The Multiband Imaging Photometer on {\it Spitzer} (MIPS, Rieke et al. 2004) observed
133P/Elst-Pizarro using 24 $\mu$m channel at three different epochs on 2005 April 11
under program 3119 (PI: W. T. Reach). The angular resolution of the MIPS camera at
24 $\mu$m band was $2.5^{\prime\prime}$ with a field of view (FOV) of
$\sim$ $5.4^\prime\times5.4^\prime$.
The integrated fluxes for 24 $\mu$m channel are measured using the method described in
\citet{Hsia14}. The aperture calibrations of this MBC at 24 $\mu$m vary in the adopted
aperture radii. We have corrected the fluxes using the aperture- and color-calibration
factors suggested by MIPS Instrument Handbook\footnote{http://irsa.ipac.caltech.edu/data/SPITZER/docs/mips/mipsinstrumenthandbook/1/}.

The photometric uncertainties of these flux measurements for 24 $\mu$m band are estimated
to be from $\sim$ 7$\%$ to 9$\%$. These values of the uncertainties are derived from the
absolute flux calibrations and standard deviations of flux determinations associated with
our aperture photometry method. The data are listed in Table \ref{obs1}.

\begin{table*}[htbp]
\renewcommand\arraystretch{1.0}
\caption{Mid-infrared observations of 133P.}
\label{obs1}
\centering
\begin{tabular}{@{}cccccccc@{}}
\hline
UT & $MA$      & $r_{\rm helio}$ & $\Delta_{\rm obs}$ & $\alpha$   & Wavelength & Flux$^f$ & Observatory\ \\
   &($^{\circ}$) & (AU)          &          (AU)    & ($^{\circ}$) & ($\mu m$)  & (mJy)& Instrument \\
\hline
2005-04-11 08:01 & -139.84 & 3.573 & 3.006 &  14.52 &  24.0     & 5.82$\pm$0.41 & Spitzer/MIPS   \\
2005-04-11 08:04 & -139.84 & 3.573 & 3.006 &  14.52 &  24.0     & 5.50$\pm$0.47 & Spitzer/MIPS   \\
2005-04-11 08:08 & -139.84 & 3.573 & 3.006 &  14.52 &  24.0     & 5.42$\pm$0.43 & Spitzer/MIPS   \\
2006-01-23 14:20 & -89.54  & 3.259 & 3.174 &  17.93 &  7.4-14.5 &     -         & Spitzer/IRS   \\
2006-01-23 14:40 & -89.54  & 3.259 & 3.174 &  17.93 & 14.0-21.7 &     -         & Spitzer/IRS   \\
2006-01-23 15:05 & -89.54  & 3.259 & 3.174 &  17.93 & 19.0-38.0 &     -         & Spitzer/IRS   \\
2010-03-17 06:21 & 175.51  & 3.662 & 3.419 &  15.67 & W3 (12.0) & 2.02$\pm$0.45 & WISE   \\
\hline
\multicolumn{8}{l}{$MA$: represents the Mean Anomaly of 133P at the time of observation.} \\
\multicolumn{8}{l}{$\alpha$: represents the angle between the vector of 133P to sun and the vector of 133P to telescope.}\\
\multicolumn{8}{l}{$f$: The Spitzer/IRS spectra contain too many data sets, so we do not list them in this table.}\\
\end{tabular}
\end{table*}

\subsubsection{Spitzer IRS Spectrum}
The mid-infrared spectra of MBC 133P/Elst-Pizarro were obtained by the {\it Spitzer
Infrared Spectrograph} (IRS; Houck et al. 2004) through the observation program 88
(PI: D. Cruikshank) with Astronomical Observation Request (AOR) key of 4870400.
The data were all obtained on 2006 January 23. The measurements were observed using
the Short-Low (SL) module (7.4 $\mu$m - 14.5 $\mu$m) and the Long-Low (LL) module
(14.0 $\mu$m - 38.0 $\mu$m) with spectral dispersions of $R\sim$ 60 - 130.
The diaphragm sizes are $3.7^{\prime\prime}\times57^\prime$ and
$10.5^{\prime\prime}\times168^\prime$ in SL and LL modules respectively.
The total integration times of IRS observation ranged from 968 to 1220 s.

Data were reduced starting with basic calibrated data (BCD) from the Spitzer Science
Center's pipeline version s18.7.0 and were run through the {\bf IRSCLEAN} program to
remove bad data points. Then the {\bf SMART} analysis package \citep{Higdon2004}
was used to extract the spectra. To improve the signal-to-noise ratio (S/N) of IRS
observations, the final SL and LL spectra were performed using the combined data.
Since the IRS spectrum with short and long wavelength ranges were observed at different
epochs, some scaling is needed for the shorter wavelength observations. We scaled the
IRS SL observations by a factor of 1.83 and were able to obtain a smooth spectrum.
The journal of IRS spectroscopic observations is summarized in Table \ref{obs1}.

\subsubsection{WISE data}
The {\it Wide-field Infrared Survey Explorer} (WISE) mission has mapped entire sky in
four bands at 3.4, 4.6, 12, and 22 $\mu$m with resolutions from $6.1^{\prime\prime}$
to $12^{\prime\prime}$. All four bands were imaged simultaneously, and the exposure
times were 7.7 s in 3.4 and 4.6 $\mu$m and 8.8 s in 12 and 22 $\mu$m. Mid-infrared
imaging observation of 133P was obtained from 12 $\mu$m band and processed with
initial calibration and reduction algorithm.

The aperture photometry for this object was performed using the same method described
in \citet{Hsia14}. We adopted the color correction on the calibrated flux for WISE 12
$\mu$m band using the color correction factor given by \citet{Wright10}. To estimate
the uncertainties in flux, the standard deviations of background-subtracted flux
measurements were adopted. If we take into account the characteristic uncertainty
of flux measurement, the flux error is estimated to be about 22$\%$ for 12 $\mu$m channel.
Details of the WISE infrared photometric results are also given in Table \ref{obs1}.

\subsection{Albedo and size from NEATM}
It is lucky that the thermal infrared observations above were all taken when 133P was
far away from its perihelion and did not show observable activity, thus it is safe for
us to use them as the thermal emission from the surface of 133P's nucleus, which can
be used to derive the albedo, size and even thermal inertia of the nucleus. However,
the orientation of 133P's rotation axis is still unclear yet, so it is not appropriate
to use the so-called thermophysical model \citep[TPM,][]{Lagerros1996a} or advanced
thermophysical model \citep[ATPM,][]{Rozitis2011} to explain these data. Nevertheless,
we can still estimate the albedo and size of the nucleus from these data via the
so-called Near-Earth Asteroid Thermal Model \citep[NEATM,][]{Harris1988}.

The nucleus of 133P may have a irregular shape, but the available data cannot resolve
the shape in detail. So here, to estimate the size of 133P, we define the effective
diameter $D_{\rm eff}$ by treating it to be spherical. Then $D_{\rm eff}$  can be
related to its geometric albedo $p_{v}$ and absolute visual magnitude $H_{v}$ via
\citep{Fowler1992}
\begin{equation}
D_{\rm eff}=\frac{1329\times 10^{-H_{v}/5}}{\sqrt{p_{v}}}~(\rm km) ~.
\label{Deff}
\end{equation}
On the other hand, the Bond albedo $A_{\rm eff,B}$ can be related to the
geometric albedo $p_{v}$ by
\begin{equation}
A_{\rm eff,B}=p_{v}q_{\rm ph}~,
\label{aeffpv}
\end{equation}
where $q_{\rm ph}$ is the phase integral that can be approximated by
\begin{equation}
q_{\rm ph}=0.290+0.684G~,
\label{qph}
\end{equation}
in which $G$ is the slope parameter in the $H, G$ magnitude system of \citet{Bowell}.
The absolute visual magnitude $H_{v}$ and slope parameter $G$ of 133P have been
measured by \citet{Hsieh2010} to be $H_{v}=15.49\pm0.05$, $G=0.04\pm0.05$, which
will be used in our fitting procedure.

\begin{figure}[htbp]
\includegraphics[scale=0.58]{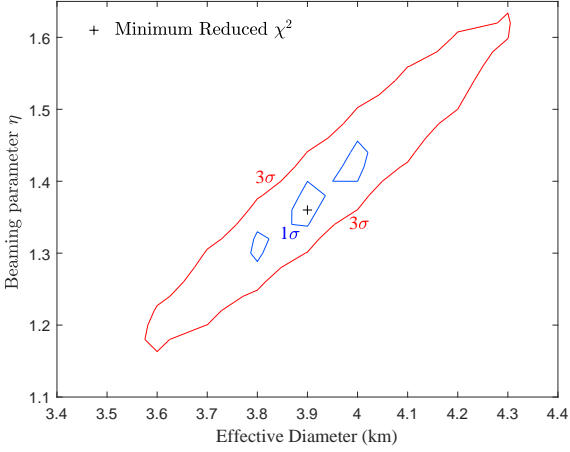}
  \centering
  \caption{NEATM fitting to the thermal-infrared date with effective diameter and beaming parameter as two free parameters. The 1$\sigma$ boundary (blue curve) corresponds to $\chi^2= 2.3$, while the 3$\sigma$ boundary (red curve) corresponds to $\chi^2=11.8$ (Press et al. 2007, p. 815).}
  \label{etaDchi2}
\end{figure}
\begin{figure}[htbp]
\includegraphics[scale=0.58]{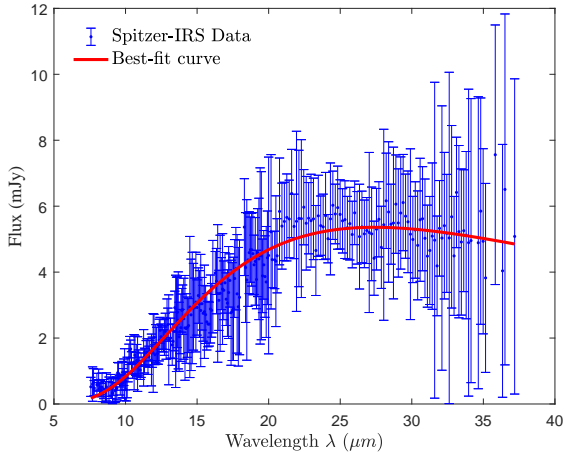}
  \centering
  \caption{Comparison between the Spitzer-IRS specta and best-fit curve by NEATM with $D_{\rm eff}=3.9$ km, $\eta=1.35$, $p_{v}=0.074$ and $A_{\rm eff,B}=0.024$.}
  \label{FluxWLs}
\end{figure}

The NEATM fitting results are presented in Figure \ref{etaDchi2}, which is a contour of the
reduced $\chi^2_{\rm r}$ with effective diameter $D_{\rm eff}$ and beaming parameter $\eta$
as two free parameters. The 1$\sigma$-level result is not that good, so we will adopt the
3$\sigma$-level results $D_{\rm eff}=3.9^{+0.4}_{-0.3}$ km, $\eta=1.35^{+0.3}_{-0.2}$.
Then the geometric albedo can be derived to be $p_{v}=0.074\pm0.013$, and the bond
albedo can be obtained as $A_{\rm eff,B}=0.024\pm0.004$, which would be useful for
thermophysical modeling. To verify the results, we plot the comparison between the Spitzer-IRS
specta and best-fit curve by NEATM in Figure \ref{FluxWLs}. The best-fit curve by NEATM matches
well to the Spitzer-IRS spectra, indicating that our radiometric results should be reliable.
We summarize the radiometric results in Table \ref{Nfresults}.
\begin{table}
\renewcommand\arraystretch{1.0}
\caption{Derived Results from NEATM fitting.}
\label{Nfresults}
\centering
\begin{tabular}{@{}lcc@{}}
\hline
Properties & $3\sigma$ Level \\
\hline
Beaming parameter     $\eta$      & $1.35^{+0.3}_{-0.2}$ \\
Effective diameter $D_{\rm eff}$  & $3.9^{+0.4}_{-0.3}$ km \\
Geometric albedo  $p_{\rm v}$     & $0.074\pm0.013$ \\
Bond albedo  $A_{\rm eff,B}$      & $0.024\pm0.004$ \\
\hline
\end{tabular}
\end{table}

\section{Temperature Constraints}
Information about temperature environment of 133P is crucial for the design of instruments
onboard the spacecraft, especially for the instruments on a lander. The temperature distribution
of a small body is largely decided by its rotation and orbital motion. But unfortunately, the
exact orientation of 133P's rotation axis is still unknown, due to the difficulty of observation
of light-curves when it is inactive. Nevertheless, the temperature environment can still be
investigated by considering various cases of orientations of the rotation axis.

\subsection{Description of Rotation Axis}
\begin{figure}[htbp]
  \includegraphics[scale=0.45]{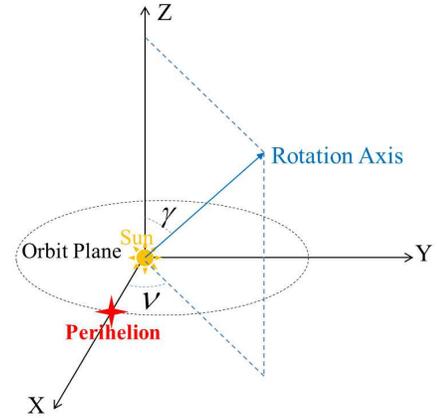}
  \centering
  \caption{Description of rotation axis of the small body with respect to its orbital plane.
  The Z-axis is the normal vector of the orbital plane, the X-axis points from the Sun to the
  perihelion of the orbit, and Y-axis is chosen to form a right-handed coordinate system with
  the Z-axis and X-axis. $\gamma$ is obliquity and $\nu$ is azimuth angle.}
  \label{RotAxis}
\end{figure}

To begin with, we need a coordinate system to give descriptions of the rotation axis.
For convenience, we introduce two parameters --- obliquity $\gamma$ and azimuth $\nu$,
to define the orientation of rotation axis with respect to the orbital plane as shown in
Figure \ref{RotAxis}.

Although the exact orientation of rotation axis of 133P is not clear yet,
\citet{Toth2006} and \citet{Hsieh2010} have obtained constraints for the
obliquity $\gamma$ according to observed light curves in 2002 and 2007.
They found that obliquity $\gamma=30\pm10^\circ$ can fit better to the
observed light curves, but the azimuth $\nu$ cannot be well constrained.

\subsection{Annual Average Temperature}
With an assumed value for the obliquity $\gamma$, the first thing that we can do
is to estimate the annual average temperature on each local latitude $\theta$ of 133P.
To do this, we need to assume infinite thermal inertia, and then the fast-rotating or
isothermal latitude model \citep{Lebofsky1990} can be applied. The annual average
temperature $\tilde{T}(\theta)$ of each latitude can be simply estimated as
\begin{equation}
(1-A_{\rm eff,B})\tilde{L}_{\rm s}(\theta)=\varepsilon\sigma \tilde{T}(\theta)^4,
\end{equation}
where $A_{\rm eff,B}$ is the bond albedo as estimated above, $\varepsilon\sim0.9$ is the
average thermal emissivity, $\tilde{L}_{\rm s}(\theta)$ is the annual average incoming
solar flux on each latitude and can be estimated via \citep{Ward1974}
\begin{eqnarray}
\begin{aligned}
\tilde{L}_{\rm s}(\theta)=
&\frac{L_{\odot}}{2\pi^2 a^2\sqrt{1-e^2}}~\times \\
&\int^{2\pi}_0\sqrt{1-(\cos\gamma\sin\theta-\sin\gamma\cos\theta\sin\varphi)^2}~{\rm d}\varphi,
\end{aligned}
\end{eqnarray}
where $L_{\odot}=1361.5~{\rm W~m^{-2}}$ is the solar constant, $a$ is the semimajor axis in AU,
$e$ is eccentricity, $\gamma$ is obliquity, $\theta$ is latitude and $\varphi$ is longitude.

\begin{figure}[htbp]
  \includegraphics[scale=0.58]{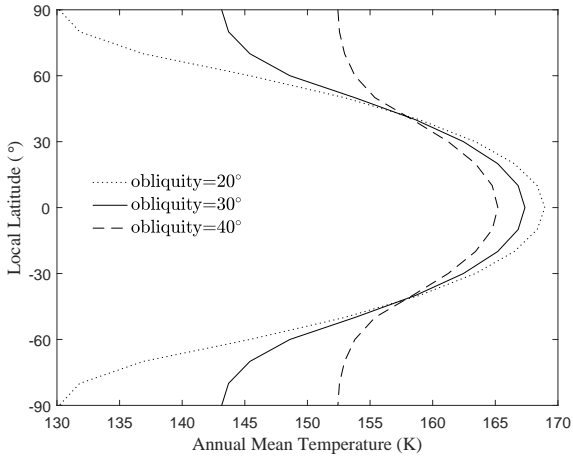}
  \centering
  \caption{Estimated annual average temperature on each local latitude of 133P on the basis of
  obliquity $\gamma=30\pm10^\circ$. The so-called local latitude $\theta$ is defined as the
  the complementary angle of the angle between the local normal vector and the rotation axis.}
  \label{AMTemp}
\end{figure}

If the rotational parameter $\gamma=30\pm10^\circ$, and with the known orbital elements
$a=3.16$ AU, $e=0.1578$ of 133P, we are able to estimate the annual average temperature
on each local latitude, as shown in Figure \ref{AMTemp}. The annual average temperature
can be about $165\sim170$ K on the equator , and be about $130\sim155$ K on the poles.

With the estimated mean temperature, we will then estimate thermal parameters of the
surface dust layer (hereafter named as dust mantle) of 133P, because thermal parameters
including thermal conductivity, specific heat capacity and thermal inertia are all
strong functions of temperature. According to \citet{Gundlach2013}, the thermal
conductivity $\kappa$ of the dust mantle on small bodies can be related to
temperature $T$, mean grain radius $b$ and porosity $\phi$ via
\begin{eqnarray}
\kappa(T,b,\phi)&=&\kappa_{\rm solid}
\left(\frac{9\pi}{4}\frac{1-\mu^{2}}{E}\frac{\gamma(T)}{b}\right)^{1/3}
\cdot f_{1}e^{f_{2}(1-\phi)}\cdot\chi
\nonumber\\
&&+8\sigma\epsilon T^{3}\frac{e_{1}\phi}{1-\phi}b ~,
\label{ktrphi}
\end{eqnarray}
where $\kappa_{\rm solid}\sim1.5\rm~Wm^{-1}K^{-1}$ is the thermal conductivity of the dust
material, $\mu$ is Poisson's ratio, $E$ is Young's modulus, $\gamma(T)$ is the specific
surface energy, $\epsilon$ is the emissivity of the material, and $f_1$, $f_2$, $\chi$ and
$e_1$ are best-fit coefficients. For more details, we refer the reader to \citet{Gundlach2013}.

While the range of grain size of surface materials may be different for various types of small
bodies (especially inactive asteroid), 133P is expected to be more cometary-like with a dust
mantle on the surface. The size frequency distribution of dust on comets is generally described
with a power law formula $f(b)\sim b^{-3.5}$, with minimum to maximum radius $b$ from 0.1 $\mu$m
to $\sim$1000 $\mu$m \citep{Rinaldi2017}. \citet{Hsieh2004,Hsieh2006,Jewitt2014} inferred
that dust particles in the observed dust tail of 133P may be mainly $\sim10~\mu m$ in radius.
So we may surmise that smaller dust grains with radius from 0.1 $\mu$m to tens of $\mu$m might
have been depleted from most part of the surface (not include newly exposed surface).
If removing dust grains $<50$ $\mu$m, dust grains with radius in $100\pm50$ $\mu$m would have
a fraction
\[\sim\int_{50}^{150}b^{-3.5}{\rm d}b~\Big/\int_{50}^{1000}b^{-3.5}{\rm d}b \sim 93.64\%\]
of the total leftover dust grains. So we assume that the mean radius of leftover
dust grains on 133P's surface may be mainly $100\pm50$ $\mu$m.

Then if considering a annual mean temperature $\sim160$ K and porosity $\phi\sim0.5$ of the
dust mantle, the mean thermal conductivity of the dust mantle can be estimated from
Equation (\ref{ktrphi}) to be $\kappa\sim1.2\times10^{-3}\rm~Wm^{-1}K^{-1}$.
If further assuming the mean grain density $\rho_{\rm d}\sim2000\rm~kg~m^{-3}$ and mean
specific heat capacity to be $c_{\rm p}\sim500\rm~Jkg^{-1}K^{-1}$, the annual mean thermal
inertia $\Gamma$ of the surface could be estimated to be
\[\Gamma=\sqrt{(1-\phi)\rho_{\rm d}c_{\rm p}\kappa}\sim25\rm~Jm^{-2}s^{-0.5}K^{-1}.\]
being close to the thermal inertia of comet nuclei, e.g. 67P \citep{Gulkis2015}.
Besides, the mean thermal diffusivity can be estimated as
\[\alpha=\frac{\kappa}{(1-\phi)\rho_{\rm d}c_{\rm p}}\sim2.4\times10^{-9}\rm~m^{-2}s^{-1},\]
and thus the seasonal thermal skin depth can be evaluated as
\[l_{\rm sst}=\sqrt{\frac{\alpha P_{\rm orb}}{2\pi}}\sim0.3\rm~m,\]
where $P_{\rm orb}$ is the orbital period of 133P. Although the estimates of these
thermophysical parameters are quite rough approximations, they are still useful for
further analysis on the thermal behaviour of the nucleus of 133P.

\subsection{Seasonal Temperature variation}
As noted above, the assumed obliquity of $\gamma\sim30^\circ$ between the rotation axis
and normal vector of the orbital plane can have significant influence on the variation
of surface temperature along the orbit. We will show that the azimuth $\nu$ defined
in Figure \ref{RotAxis} can also have significant influence on the seasonal temperature
variation.

\begin{figure*}[htbp]
  \includegraphics[scale=0.51]{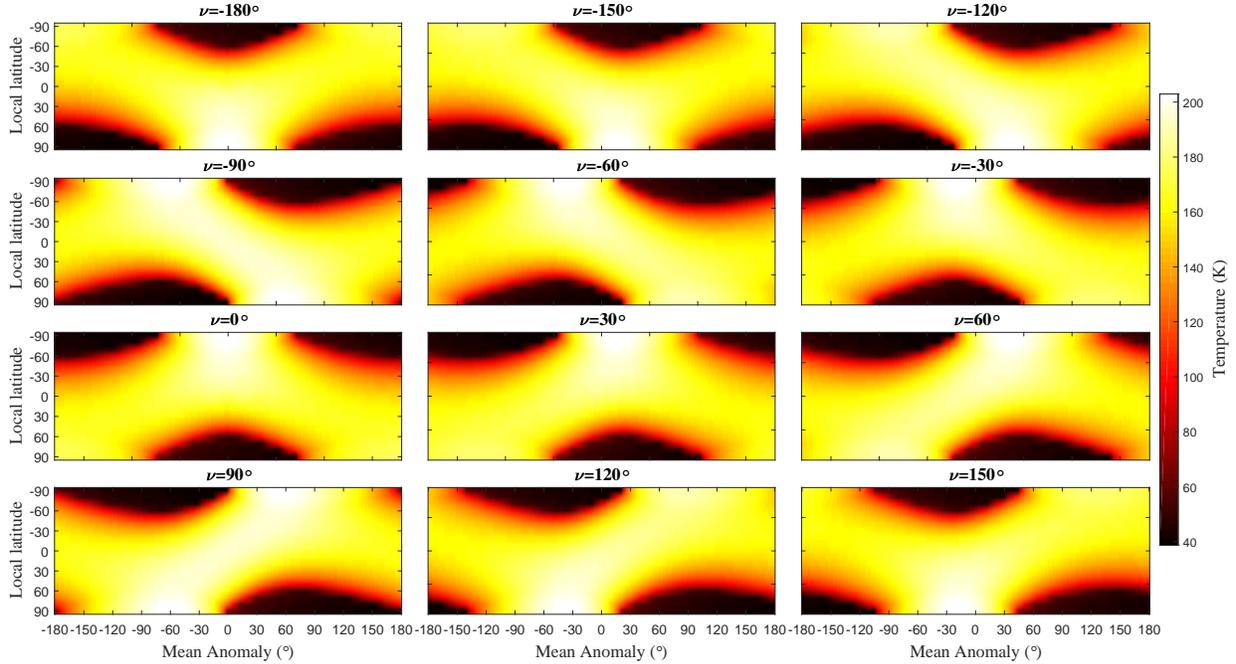}
  \centering
  \caption{Each panel represents seasonal variation of the diurnally averaged surface temperature as a function of local latitude under assumption of different azimuth $\nu$ of the spin orientation but the same obliquity $\gamma\sim30^\circ$. For different spin orientation, temperature on different latitude will get maximum (summer) or minimum (winter) at different orbital position, as the result of seasonal effect.}
  \label{Tss}
\end{figure*}
Since the azimuth $\nu$ of 133P is still unclear yet, we consider its value to vary from
-180$^\circ$ to 180$^\circ$ with step $30^\circ$. The simulated results are presented in
Figure \ref{Tss}. In each panel of Figure \ref{Tss}, the horizontal axis represents orbital
mean anomaly, the vertical axis is local latitude, and the color index stands for diurnally
averaged surface temperature. Each panel is obtained under assumption of different azimuth
$\nu$ of the spin orientation but the same obliquity $\gamma\sim30^\circ$.

According to Figure \ref{Tss}, we can clearly see that temperature on each local latitude
can reach maximum (summer) or minimum (winter) at different orbital position as a result of
seasonal effect. Temperature on the poles can vary from $\sim40$ K to $\sim200$ K.
Such seasonal variation can cause similar variation of gas/dust production if there exist
near-surface ice. The distribution of ice on 133P can be investigated if we have enough
observations on the activity of 133P. In the following section, we will present the available
observations at present on the activity of 133P, and what we can learn from these observations.

\section{Activity Constraints}
\subsection{Available observations}
\begin{figure*}[htbp]
  \includegraphics[scale=0.71]{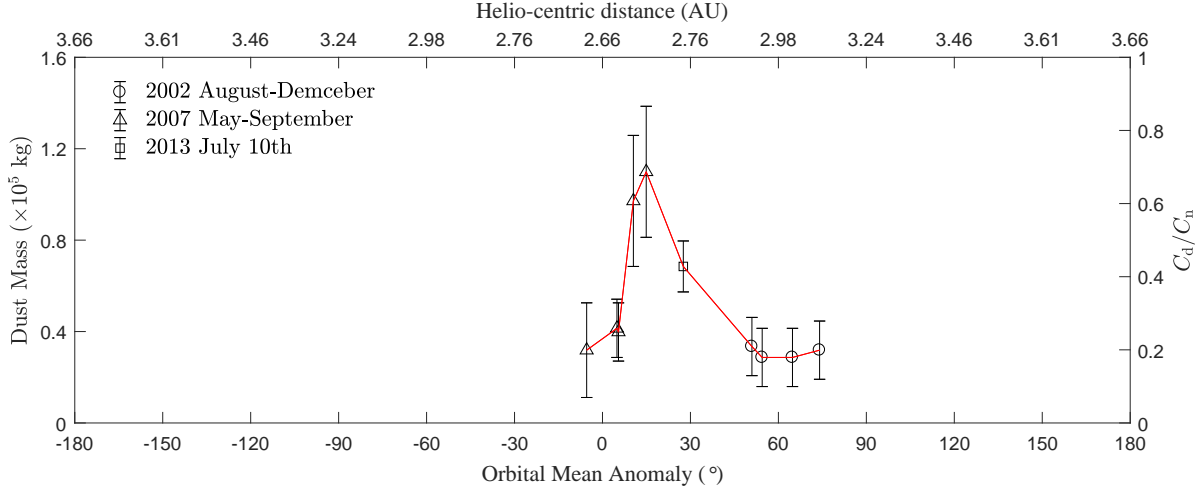}
  \centering
  \caption{Vertical axis: The ratio of light-scattering area of dust to that of nucleus (right side: $C_{\rm d}/C_{\rm n}$) and estimated total dust mass (left side: $M_{\rm dust}$) based on data in Table \ref{obs2}. The horizontal axis represents orbital mean anomaly. The variation of the total dust mass shows significant seasonal variation. The slope of the dust-mass variation curve can give us constraints on the dust production rate.}
  \label{obsCAdust}
\end{figure*}

\cite{Hsieh2004,Hsieh2010,Jewitt2014,Snodgrass2018} reported optical photometry of 133P along
its orbit, showing that activity of dust tail can appear between Mean Anomaly $\sim -5.4^\circ$
and $\sim 74^\circ$. \cite{Hsieh2010,Jewitt2014} also measured the ratio of light-scattering
area of dust $C_{\rm d}$ to that of nucleus $C_{\rm n}$ according to the photometric images.
In the above section, we have computed the effective diameter $D_{\rm eff}$ of 133P to be
about $3.9$ km. So if we assume that the dust particles in the tails have similar albedo to
that of the nucleus, and have an average radius $b_{\rm d}\sim10~\mu m$ \citep{Hsieh2004},
we could estimate the total dust mass via
\begin{equation}
M_{\rm dust}=\frac{1}{3}\pi\rho_{\rm d}b_{\rm d}\frac{C_{\rm d}}{C_{\rm n}}D_{\rm eff}^2,
\end{equation}
where $\rho_{\rm d}$ represents the mass density of dust particles. The values of
$C_{\rm d}/C_{\rm n}$ and $M_{\rm dust}$ are listed in Table \ref{obs2}, and plotted in
Figure \ref{obsCAdust} as functions of orbital mean anomaly. The variation of the produced
dust mass show significant seasonal variation, which could provide us estimation on the
dust production rate, and even constraints on the distribution of ice on 133P.
\begin{table}
\centering
\renewcommand\arraystretch{1.0}
\caption{Previous photometry observations of 133P \citep{Hsieh2004,Hsieh2010,Jewitt2014}.}
\label{obs2}
\begin{tabular}{@{}cccc@{}}
\hline
UT & $MA$        & $C_{\rm d}/C_{\rm n}$  & $M_{\rm dust}$ \\
   &($^{\circ}$) &                        &   ($10^5$ kg)  \\
\hline
2007-05-19  & -5.4  & 0.20$\pm$0.13  & 0.3186$\pm$0.2072 \\
2007-07-17  &  4.9  & 0.26$\pm$0.08  & 0.4141$\pm$0.1274 \\
2007-07-20  &  5.5  & 0.25$\pm$0.08  & 0.3982$\pm$0.1274 \\
2007-08-18  & 10.5  & 0.61$\pm$0.18  & 0.9716$\pm$0.2867 \\
2007-09-12  & 14.9  & 0.69$\pm$0.18  & 1.0990$\pm$0.2867 \\
2013-07-10  & 27.6  & 0.43$\pm$0.07  & 0.6849$\pm$0.1115 \\
2002-08-19  & 51.0  & 0.21$\pm$0.08  & 0.3345$\pm$0.1274 \\
2002-09-08  & 54.5  & 0.18$\pm$0.08  & 0.2867$\pm$0.1274 \\
2002-11-06  & 64.8  & 0.18$\pm$0.08  & 0.2867$\pm$0.1274 \\
2002-12-28  & 74.1  & 0.20$\pm$0.08  & 0.3186$\pm$0.1274 \\
2008-10-27  & 86.8  & Faint dust     & -\\
\hline
\multicolumn{4}{l}{$MA$: represents the Mean Anomaly of 133P.} \\
\multicolumn{4}{l}{$C_{\rm d}$: light scattering area of dust.} \\
\multicolumn{4}{l}{$C_{\rm n}$: light scattering area of nucleus.} \\
\multicolumn{4}{l}{$M_{\rm dust}$: Estimated total dust mass if dust radius $\sim10~\mu m$.} \\
\end{tabular}
\end{table}

\subsection{Dust/gas production rate}
The slope of the dust-mass variation curve in Figure \ref{obsCAdust} indicates the production
rates of dust, which also varies with the orbital position. We can infer that activity at least
starts at around mean anomaly $-5.4^\circ$, where the slope of dust mass indicates total dust
production $\sim0.0017\rm~kgs^{-1}$, and total water gas production rate $1.1\times10^{22}\rm~s^{-1}$
if assuming dust-ice mass ratio $\sim5:1$ similar to that of 67P \citep{Fuller2016}. We also
find that the slop seems to get maximum at around $8\pm3^\circ$, indicating that 133P may be
most active during this time range with total dust production rate $\sim0.0215\rm~kgs^{-1}$,
and total water gas production rate $1.4\times10^{23}\rm~s^{-1}$.

The estimation for the maximum water production rate here is consistent with the upper
limit of water production rate $<\sim10^{24}\rm~s^{-1}$ given in \citet{Licandro2011}.
But such water production rate is far weaker than that $\sim10^{26}\rm~s^{-1}$ of a typical
JFC like 67P at similar helio-centric distance $\sim2.7$ AU \citep{Hansen2016}, indicating
the existence of a dust mantle on the surface, thus lowering down the gas production rate.
But the question about how and where the gas are produced from the nucleus of 133P, namely,
whether the gas is produced by sublimation from homogeneous buried ice layer or only from
regional near-surface ice patches, is still unsolved. We will discuss such question in the
following sections.

\subsubsection{Homogeneous buried ice layer?}
\begin{figure}[htbp]
\includegraphics[scale=0.58]{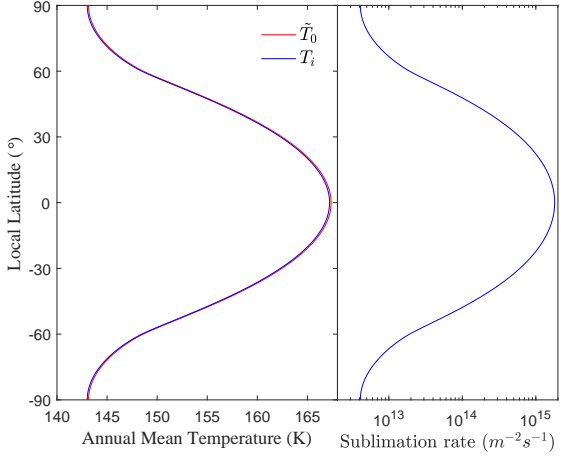}
  \centering
  \caption{Left panel: Equilibrium subsurface temperature $\tilde{T}_0$ of the dust mantle and
  the ice front temperature $T_i$ at each local latitude, in the case of $\tilde{T}_0<180$ K,
  $T_i$ nearly equals to $\tilde{T}_0$. Right panel: sublimation rate of water ice below each latitude.
  }\label{TsTiJn}
\end{figure}

If it is assumed that 133P has a homogeneous two-layer system with a dust mantle covering a
dust-ice mixture interior. The thickness of the dust mantle should be $\sim 50$ to $150$ m
if 133P has stayed in the main belt over the entire lifetime of the Solar System according
to \cite{Prialnik2009}. For such a two layer system, the "two layer sublimation model" developed
in \cite{Yu2019} can be well applicable. But if 133P is a newly formed fragment of a larger
icy parent object, the ice layer can be closer to the surface, and the dust mantle can be
much thinner. The question is how thin the dust mantle could be in the case of a homogeneous
distribution? If we expect the existence of a stable dust mantle on 133P, the dust mantle thickness
is then expected to be several seasonal thermal skin depths, like $2l_{\rm sst}\sim 0.6\rm~m$.
Then the "two layer sublimation model" \citep{Yu2019} can still be a good approximation. If adopting
the obliquity of 133P to be about $\pm30^\circ$, then the seasonal equilibrium subsurface temperature
$\tilde{T}_0$ and corresponding ice sublimation front temperature $T_i$ at each local latitude
can be estimated, as shown in left panel of Figure \ref{TsTiJn}.

The right panel of Figure \ref{TsTiJn} presents model estimated water sublimation rate
of the ice front below each local latitude from $4\times10^{12}\rm~m^{-2}s^{-1}$ to
$3\times10^{15}\rm~m^{-2}s^{-1}$. The total water gas production rate of such homogeneous
case can be estimated to be $\sim10^{22-23}\rm~s^{-1}$, which is quite close to
the above estimated production rate from observation. Thus we may say that the
homogeneous case is reasonable if just considering the total production rate
$<\sim10^{23}\rm~s^{-1}$.

However, in order to be able to eject dust particles, the drag force of the outflow
gas has to overcome the gravity force of the dust particles
\begin{equation}
F_{\rm drag}=C_{\rm d}\pi b_{\rm d}^2\hat{m}\tilde{v}_{\rm th}J_{\rm s} >
\frac{4}{3}\pi b_{\rm d}^3\rho_{\rm d} g_{\rm s},
\end{equation}
where $\tilde{v}_{\rm th}$ represents the mean thermal velocity of water molecular at surface,
$J_{\rm s}$ means the outflow number flux at surface and
\[g_{\rm s}\sim \frac{GM}{R^2}-R\omega^2\cos^2\theta\]
stands for gravitational acceleration at the surface of 133P. As noted above, dust particles
in the dust tail are mainly $10\rm~\mu m$ in radius \citep{Hsieh2004,Hsieh2010}, the outflow
water flux is hence required to be $J_{\rm s}>\sim5\times10^{17}\rm~m^{-2}s^{-1}$,
indicating a sublimation temperature $T_{\rm i}>\sim 185 \rm K$. So the maximum sublimation
rate $\sim3\times10^{15}\rm~m^{-2}s^{-1}$ of the homogeneous case shown in Figure \ref{TsTiJn}
is too weak to drag away dust particles with radius $10\rm~\mu m$ from the nucleus surface
of 133P, although such homogeneous case can have similar total water production rate to the
observation constraint.

On the other hand, from considering the maximum total water gas production rate $\sim10^{23}\rm~s^{-1}$,
and the requirement of sublimation rate $>\sim5\times10^{17}\rm~m^{-2}s^{-1}$, we can estimate that
the sublimation area should be $<\sim3\times10^5\rm~m^2$, corresponding to a circular area with
diameter $<\sim 600$ m, which is a small region on the surface (nearly $<1/10$ of the total surface
of 133P). Therefore, we tend to believe that sublimation of homogeneous buried ice layer
is unlikely to be responsible for the observed activity, and instead regional near-surface
ice patch is necessary to explain the observations.

\subsubsection{Regional near-surface ice patch?}
So if we expect a near-surface ice patch to be the explanation of the observations, seasonal
temperature variation of the ice patch should be responsible for the observed seasonal
variation of dust tail activity. The dust activity was observed to start at around mean
anomaly $-5.4^\circ$, indicating sublimation temperature $\sim180\pm5$ K at this orbital
position. The estimated total water production rate at around mean anomaly $8\pm3^\circ$
is nearly 10 times larger than that at mean anomaly $-5.4^\circ$. If assuming the sublimation
area of the ice patch to be unchanged, the sublimation rate at mean anomaly $8\pm3^\circ$ would
also be 10 times larger, thus giving $J_{\rm s}>\sim5\times10^{18}\rm~m^{-2}s^{-1}$, and the
sublimation temperature to be around $200\pm10$ K.
Then the area of the near surface ice patch can be estimated to be $<\sim3\times10^4\rm~m^2$,
corresponding to a circular area with diameter $<\sim 200$ m, which is indeed a small
region on the surface of 133P. The results for the production rates and sublimation
temperature are summarized in Table \ref{dpr}.
%\footnotesize
\begin{table}
\centering
\renewcommand\arraystretch{1.0}
\caption{Estimated dust and water gas production rate of 133P.}
\label{dpr}
\begin{tabular}{@{}lcccc@{}}
\hline
  & $MA$        & Dust        & Water Gas  & Temperature\\
  &($^{\circ}$) & (kgs$^{-1}$)  & (s$^{-1}$) &  (K) \\
\hline
Start     & -5.4      & 0.0017  &  $1.1\times10^{22}$ & $185\pm5$ \\
Maximum   &  8$\pm$3  & 0.0215  &  $1.4\times10^{23}$ & $200\pm10$ \\
Stop      &  74.1     & -       &  - & \\
\hline
\multicolumn{5}{l}{$*$: Assuming dust to gas mass ratio $\sim$5:1.} \\
\end{tabular}
\end{table}

If assuming that the possible ice patch is in the bottom of one bowl-shaped crater, the diameter
of the crater-rim has to be on the order of $\sim 200$ m. Such small crater can be created from
one impact event by an impactor with diameter $\sim$ 20 m and impact velocity $\sim$ 10 km/s
\citep{Vincent2015}. The estimated diameter of the crater-rim can further tell us that the depth
of the crater should be around $\sim40$ m according to the nearly 5:1 ratio of crater-rim-diameter
to crater-depth given in \citet{Pike1974}. This scenario indicates that the internal ice layer
should be buried below $\sim40$ m depth from the surface before the impact event. If 133P was
initially composed of a homogeneous mixture of dust and water ice, the time scale to form a
dust mantle with thickness $h_{\rm i}\sim40$ m in its current orbit can be estimated as
\begin{equation}
t_{\rm m}\sim\frac{h_{\rm i}^2}{2R_{\rm r}}
\end{equation}
via the long-term retreating model of buried ice layer described in \citet{Yu2019}, where
\begin{equation}
R_{\rm r}=\frac{\hat{m}\tilde{\beta} n_{\rm E}}{(1-\phi)\rho_{\rm d}\chi_0},
\end{equation}
is defined as the 'retreating rate' of buried ice layer, in which $\hat{m}$ is the mass of
water molecular, $\tilde{\beta}$ is the mean Knudsen diffusion coefficient, $n_{\rm E}$ is
the saturation number density under the temperature $T_{\rm i}$ of the buried sublimation
front, and $\chi_0\sim0.15$ is the ice/dust mass ratio. For the current 133P with a mean
temperature $T_{\rm i}<\sim165$ K of the buried sublimation front,
\[R_{\rm r}<\sim2.6\times10^{-13}\rm~m^2s^{-1},\]
thus giving
\[t_{\rm m}>\sim100\rm~Myr.\]
Therefore, it can be expected that 133P has been in the main-belt for a long time, with
the ice layer deeply buried below $\sim40$ m depth in most regions. This time scale is
much larger than the proposed age $\leq14$ Myr \citep{Carruba2019} of the young Beagle family
that is thought to associate with 133P \citep{Nesvorny2008}, indicating that 133P is more
likely to be a relatively old planetesimals or a member of an old family (e.g. Themis family)
than a recently formed fragment of a young asteroid family.

If it is assumed that the buried ice layer has similar dust/ice mass ratio and dust size distribution
as those of fresh JFCs, then sublimation of the exposed ice patch can be active enough to blow away
$\sim10~\mu$m dust particles and hence generate dust tails like the observations. During the early
tens of orbital cycles, the strong sublimation of the ice patch can blow away most dust particle.
A very thin dust mantle with thickness $<1$ cm may form on the proposed ice patch just like
the surface layers of 67P. But such a thin dust mantle would be unstable and can be repeatedly formed
and destroyed following the diurnal or orbital cycles, causing a moving boundary and hence preventing
the formation of a stable dust mantle until the accumulation of a sufficient amount of large dust
particles ($>2$ cm) on the surface. In this way, the current observation might be explained.

\subsection{Possible location of the exposed ice patch}
\begin{figure}[htbp]
\includegraphics[scale=0.58]{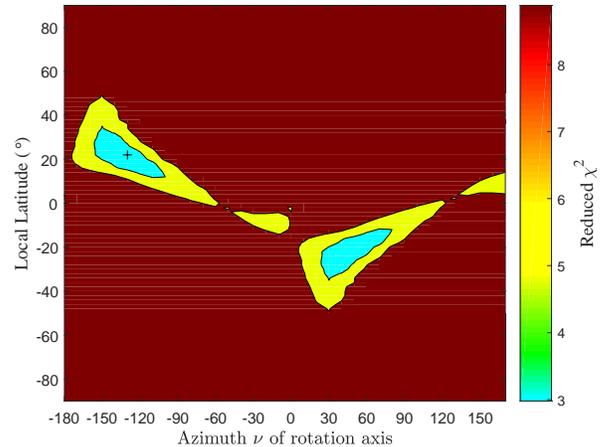}
  \centering
  \caption{Contour of reduced $\chi^2$ based on fitting to Table \ref{dpr} with two free parameters --- location latitude $LA_{\rm i}$ of the possible icy patch and the azimuth $\nu$ of rotation axis. The black '+' stands for the best-fit
  result, the cyan region stands for 1$\sigma$-level constraint and the yellow region means 3$\sigma$-level constraint.
  }\label{fitChi2}
\end{figure}

Now the question of most interest is where the exposed icy patch can be located on the surface?
The significant seasonal variation of activity as shown in Figure \ref{obsCAdust} should be
results of time variation of temperature of the near-surface ice patch, thus giving constraints
for the sublimation temperature as shown in Table \ref{dpr}. Moreover, temperature on different
local latitudes can show different seasonal variations as the result of some particular orientation
of rotation axis with respect to the orbit (as shown in Figure \ref{Tss}). This relation would
provide a way to investigate the possible location of the surface ice patch and orientation of
the rotation axis. We treat the location latitude $LA_{\rm i}$ of the possible ice patch and
the azimuth $\nu$ of rotation-axis as two free parameters to fit the previously obtained values
for the sublimation rate and sublimation temperature in Table \ref{dpr}. The fitting results
are presented in Figure \ref{fitChi2}.

Figure \ref{fitChi2} shows the contour of reduced $\chi^2$ obtained by fitting $LA_{\rm i}$
and $\nu$ to sublimation temperature given in Table \ref{dpr}. The location of the lowest
reduced $\chi^2$ indicates the best fit to be $LA_{\rm i}\sim20^\circ$ and $\nu\sim-130^\circ$.
The cyan region stands for 1$\sigma$-level constraint for $LA_{\rm i}$ and $\nu$, giving
$LA_{\rm i}=10\sim40^\circ$, $\nu=-160\sim-100^\circ$ or $LA_{\rm i}=-40\sim-10^\circ$,
$\nu=20\sim80^\circ$. The yellow region means 3$\sigma$-level constraint, giving
$LA_{\rm i}=-50\sim50^\circ$ only.

\begin{figure*}[htbp]
\includegraphics[scale=0.58]{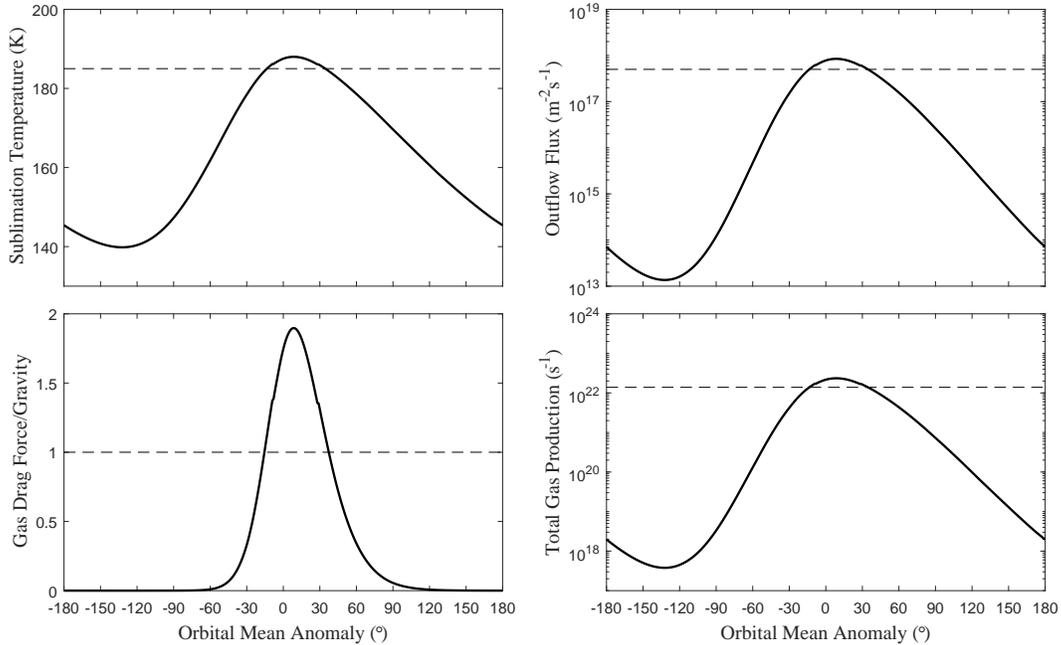}
  \centering
  \caption{Simulated seasonal variation of sublimation temperature, outflow flux, ratio of gas drag force to gravity force on  $10~\mu$m dust particles, and total water production rate assuming a regional near-surface ice patch with diameter $\sim200$ m, locating at latitude $\sim20^\circ$ of the nucleus that has rotation axis $\gamma\sim30^\circ$ and $\nu\sim-130^\circ$ with respect to the orbital plane of 133P. The horizontal dashed line in each panel represents the critical value when $10~\mu$m dust particles can be ejected.
  }\label{TiJsDgr}
\end{figure*}

In terms of 3$\sigma$-level contour, the azimuth $\nu$ still cannot be well constrained
due to the lack of enough information for seasonal variation of sublimation rate on 133P
from observations. Nevertheless, we can at least infer the location latitude $LA_{\rm i}$
of the possible ice patch to be between latitude $-50\sim50^\circ$, indicating that the possible
ice patch is unlikely to be located at high latitudes $>50^\circ$. If using the best-fit result
$LA_{\rm i}\sim20^\circ$ and $\nu\sim130^\circ$ as an example, we can simulate how
the sublimation temperature, sublimation rate and hence gas drag force, total water production
rate of the near-surface ice patch vary with orbital movement. The results are presented in
Figure \ref{TiJsDgr}.

Figure \ref{TiJsDgr} clearly shows how the seasonal variation of sublimation temperature of a
regional near-surface ice patch affects the sublimation rate of water and hence the ejection
of dust. In such a case, the ratio of gas drag force to gravity force on $10\mu$m dust particles
will only be larger than 1 when 133P gets close to perihelion within a short time interval, thus
only producing dust tail during this period. But the total water gas production rate is too
low to from an observable gas coma and dust coma, which may be the reason why we only observed
long narrow dust tails behind 133P.

\section{Discussion and Conclusion}
Although 133P is famous for being the first recognized main-belt comet and has been discovered
for $\sim$40 years, we still know very little about physical properties of this object, which is
rather disadvantageous if we need to plan space mission to study it up-close. This paper therefore aims
to obtain estimates for the basic physical parameters of 133P, including size, albedo, temperature
and even activity mechanism. Firstly, by NEATM fitting to the data from Spitzer-MIPS, Spitzer-IRS
and WISE, we obtain estimates for the effective diameter $D_{\rm eff}=3.9^{+0.4}_{-0.3}$ km,
geometric albedo $p_{\rm v}=0.074\pm0.013$ and mean Bond albedo $A_{\rm eff,B}=0.024\pm0.004$.
The derived diameter is close to the result of \citet{Hsieh2009}, which used a similar NEATM
procedure but only fitted the Spitzer-MIPS data by assuming a beaming parameter $\eta\sim1.0$.
The estimated $p_{v}$ is closer to the result of \citet{Bagnulo2010}, which utilized a
different method based on polarization measurement when 133P was active. The advantages of our
results in comparison to previous results of \citet{Hsieh2009} and \citet{Bagnulo2010} mainly
lie in two aspects: First, the mid-infrared data were all obtained when 133P was far away from
perihelion and did not show observable activity, these data are consequently thermal emission
from the nucleus of 133P completely without pollution by dust activity; Second, the data cover
three different epochs, namely three different solar phase angles, making it possible to remove
the degeneracy of diameter and beaming parameter in the NEATM procedure, and hence simultaneously
constrain the diameter (albedo) and beaming parameter.

Of course, the method NEATM we use naturally bears disadvantages that the effects of thermal
inertia and rotation axis are not well resolved, which could influence the estimates for size
and albedo. However, currently the rotation axis of 133P is unclear yet, making it difficult
to use the so-called thermophysical model (TPM) to derive size, albedo and thermal inertia
simultaneously. In such a condition, the more reasonable way is: firstly using NEATM to
compute size and albedo, secondly using albedo to estimate mean temperature, finally using
mean temperature to estimate thermal parameters.

Actually, it is unavoidably that we still don't know the rotation axis of 133P, because there
are not enough light-curves of 133P to do light-curve inversion procedure. Light-curve observation
of 133P is too difficult for small optical telescope due to the far distance and small size of 133P,
and it is quite difficult to apply large telescopes to observe 133P. Thus we need other ways to
investigate the rotation axis of 133P. As what we have done in this work, the seasonal variation
of activity could provide us a way to investigate the rotation axis.

Since the production rate estimated from current observations is too low in comparison to
that of typical JFC like 67P, the activity of 133P is unlikely to be caused by sublimation
of homogeneous buried ice layer. We thus believe that the activity of 133P might have been
generated by the sublimation of a regional near-surface ice patch. The estimated diameter
$<\sim200$ m of the proposed ice patch can be generated from one impact event by an impactor
with diameter $\sim$ 20 m and impact velocity $\sim$ 10 km/s \citep{Vincent2015}. We know
that current 133P can show activity when it is at mean anomaly $-5.4^\circ\sim74.1^\circ$. But
before August 7, 1996, observations of 133P on July 24, 1979 (mean anomaly $\sim15.3^\circ$),
and on September 15, 1985 (mean anomaly $\sim48.4^\circ$) did not show activity. Thus it is
possible that the proposed ice patch might have been exposed by one impact event during
year 1985-1996.

The seasonal feature that dust activity only appears close to or shortly after perihelion
further supports the idea of regional ice patch. Then the location of the ice patch becomes
another unknown problem besides rotation axis, which together decide the seasonal variation
of 133P's activity, providing us a way to investigate the location of the ice patch and
the orientation of rotation axis. Based on current activity observations, the 3$\sigma$-level
constraint for the rotation axis is not good yet, making the solution of ice-patch location
non-unique as well. However, if we get sufficient observations on the activity of 133P to
describe the seasonal variation of dust or gas production rate in future, we are sure that the
rotation axis of the nucleus as well as the location of the possible near-surface ice patch
could be well approximated by this way.

In conclusion, we find that the main-belt comet 133P is largely different from typical JFCs,
not only orbital features but also distribution of ice in the nucleus. The current activity of
133P might be re-triggered by one impact event during year 1985-1996, before which 133P may be
an extinct comet (or ice-rich asteroid) with ice layer buried below $\sim40$ m depth from the
surface. The time scale to form such a thick dust mantle by sublimation loss of water is estimated
to be $>100$ Myr, being much larger than the age of the young Beagle family, indicating that 133P
is more likely to be formed from an old family than a youn one, or probably a relatively old
planetesimals survived from the dawn of the solar system. The proposed impact event may expose
a regional near-surface ice patch with diameter $<\sim200$ m, probably locating somewhere between
latitude $-50\sim50^\circ$. The seasonal variation of temperature of the exposed ice patch will
thus generate the seasonal feature of activity as shown by current observations.

\section*{Acknowledgments}
We would like to thank Professor Dina Prialnik and Professor Henry Hsieh for discussions to improve this work, and thank the WISE and Spitzer teams for providing public data. This work was supported by the grants from The Science and Technology Development Fund, Macau SAR (file no. 119/2017/A2, 061/2017/A2 and 0007/2019/A) and Faculty Research Grants of The Macau University of Science and Technology (program no.FRG-19-004-SSI).

\bibliographystyle{named}


%\bibliography{sample63}{}
%\bibliographystyle{aasjournal}



\end{document}